\documentclass[12pt, letterpaper]{article}

\usepackage{amsmath,amssymb}
\usepackage{cite}
\usepackage{fancyhdr}
\usepackage[top=1in, bottom=1in, left=1in, right=1in]{geometry}
\usepackage[dvipdfm, dvips]{graphicx}
\usepackage{hyperref}

\numberwithin{equation}{section}

\begin{document}


\setcounter{page}{0}
\date{}

\lhead{}\chead{}\rhead{\footnotesize{RUNHETC-2012-07\\SCIPP-12/06}}\lfoot{}\cfoot{}\rfoot{}

\title{\textbf{Arguments Against a Finite ${\cal N} = 8$ Supergravity\vspace{0.4cm}}}

\author{Tom Banks$^{1,2}$ \vspace{0.7cm}\\
{\normalsize{$^1$NHETC and Department of Physics and Astronomy, Rutgers University,}}\\
{\normalsize{Piscataway, NJ 08854-8019, USA}}\vspace{0.2cm}\\
{\normalsize{$^2$SCIPP and Department of Physics, University of California,}}\\
{\normalsize{Santa Cruz, CA 95064-1077, USA}}}

\maketitle
\thispagestyle{fancy}

\begin{abstract}
\normalsize \noindent
I present arguments based on dispersion theory and the existence of charged BPS states of small charge, that maximally supersymmetric SUGRA
is not a finite theory.  The essence of the argument is that the perturbative
construction of the theory implies that scattering amplitudes are invariant under a continuous $SU(8)$ global symmetry, which is broken by contributions from virtual BPS states. Conventional estimates of the contributions of those states above production threshold, shows that they should be order one. If the amplitude satisfies a dispersion relation with a finite number of subtractions, as has recently been argued by Giddings and Porto\cite{giddport}, then $SU(8)$ violating amplitudes appear at finite orders in the low energy expansion. We also make a brief remark about the inconsistency of the idea of {\it classicalization} in non-gravitational theories.
\end{abstract}


\newpage
\tableofcontents
\vspace{1cm}

\section{Introduction}

There has been a lot of interest in recent years in the question of whether $N=8$ supergravity (SUGRA) has a finite S-matrix in perturbation theory\cite{interest}.   Most of the debate has centered around detailed perturbative calculations and the question of whether the finite answers that have been obtained could be explained in terms of symmetries, or were surprises that might point to new principles at work, which could guarantee finiteness to all orders.  The work of \cite{elvangetal} shows that all existing calculations can be explained by symmetries, and that the first real test of finiteness will come at 7 loops.  I believe that this is now the consensus in the field.

I've always been skeptical about the finiteness arguments, for reasons that I will outline below.  At the Penn State Conference in the Fall of 2009 I gave a talk claiming that the putative finite theory could not be a real theory of gravitation, because it could not contain the BPS black hole states that can be found as solutions of the classical equations of $N=8$ SUGRA.  In that talk, in order to make a comparison with the way in which ``non-perturbative" states affect string perturbation theory, I emphasized the black hole states with $E(7,7)$ invariant $J_4 = 0$, which go to zero mass at the boundaries of moduli space.  In the string theoretic realizations of maximal SUGRA, these are elementary particles in the appropriate limits, and have very small entropy.
The entropy is of course an invariant on the moduli space, and, as a consequence, the corresponding classical solutions have zero area horizons and are singular.

In \cite{bfk} it was suggested that a version of the theory omitting only these singular black holes might be consistent.   While I disagree with this claim, it seems easier to avoid the $J_4 = 0$ black holes, because the argument does not depend on them.   Instead, it depends only on the following basic points

\begin{itemize}

\item Theories of Quantum gravity do not  have global continuous symmetries.  This is basically a consequence of the existence of black holes and Hawking radiation.  A global continuous symmetry  implies that we can make black holes with size of order the Planck scale, and masses of order the Planck mass, with arbitrarily large numbers of quantum states, corresponding to high dimensional representations of the group.  This violates the Covariant Entropy Bound\cite{fsb}, and it also leads to a variety of paradoxes\cite{susskind}.  In particular, the $SU(8)$ global symmetry of $N=8$ SUGRA must be broken. The methods that are used to compute finite $N=8$ loop amplitudes would guarantee that to all orders in the loop expansion, $SU(8)$ is a valid symmetry.

\item As a consequence of the Dirac quantization condition, the spectrum of BPS black holes is {\it not} $SU(8)$ invariant.  Using unitarity, we can then prove that amplitudes above the threshold for black hole production for scattering of particles in the SUGRA multiplet, also violate $SU(8)$.  $SU(8)$ invariance follows from SUSY for four point amplitudes, so the simplest such amplitude is a five point function.  An example is the $2 \rightarrow 3$ process, in which two gravitons scatter to produce two gravitons and a scalar.  The gravitons are singlets, and the scalar is in the $70$ of $SU(8)$.
This amplitude is known to be non-vanishing in perturbative string theory\cite{helvkeir}, where perturbative winding and Kaluza Klein modes are known to be dual to BPS black holes. \cite{helvkeir} shows that even tree level string theory breaks the $SU(8)$ symmetry down to $SU(4) \times SU(4)$.  Loops of BPS states break it to a discrete group.   Green, Ooguri and Schwarz\cite{gos} have shown that there are no limits of string theory in which all the BPS states decouple.

\item One often hears the phrase, ``black holes are non-perturbative states and their production will not appear in perturbation theory".  This is a mistake.
The phrase is only correct for black holes, like those of large charge, whose horizon radius is much larger than the Planck scale.  We argue that states of small charge are produced with amplitudes of order $1$ above threshold.  The thresholds are of order the Planck scale and so is the physical size of the state (by extrapolation from the classical solutions).  We recall that the production cross sections for solitons are suppressed exponentially {\it in the ratio of their physical size to their Compton wavelength}.  There is no such suppression for black holes of small charge, whether or not their invariant entropy vanishes.     We also argue that the existence of the small charge black holes follows from the existence of classical solutions plus cluster decomposition\footnote{The last argument is due to Andy Strominger.}.

\item  The rest of the argument follows if the $SU(8)$ violating $5$ point amplitude satisfies a dispersion relation in the center of mass energy, with a finite number of subtractions, when the other kinematic invariants are fixed. Such a representation shows that the amplitude has a finite perturbation series in powers of the energy over the Planck mass, which contradicts the claim that there is an $SU(8)$ invariant finite power series, constructed by the methods of \cite{interest}\footnote{ Although there are many alternative methods of calculation in the papers of reference \cite{interest}, all but the string theory based method begin with manifestly $SU(8)$ invariant tree level amplitudes and use unitarity and analyticity to construct the loops.  Interestingly, the methods of Bern, Dixon and collaborators use dimensional regularization to deal with infrared divergences. Zvi Bern has informed me that this breaks the manifest $SU(8)$ symmetry, though the final results are $SU(8)$ invariant in four dimensions. One would imagine that, since IR divergences come only from soft graviton emission, and gravitons are $SU(8)$ singlets,  one is guaranteed to get invariant results unless genuine UV divergences are encountered.}, which coincides with the amplitudes that include contributions from BPS states.

  There are two reasons to worry about the existence of such a dispersion relation.  At fixed impact parameter, we expect graviton scattering at sufficiently high energy to be dominated by black hole production (these are neutral black holes).  As a consequence of the thermal nature of black hole evaporation, any exclusive channel with a small number of high energy particles, is exponentially suppressed.   This means that, at fixed impact parameter, the amplitude cannot be power law bounded in all directions in the complex energy plane, because it violates the Cerulus-Martin bound\cite{cerulusmart}.  However, Giddings and Porto have recently argued that at {\it fixed kinematic invariants}, the shadow scattering of the black hole absorption process, produces an amplitude that {\it does} satisfy the CM bound.  

The other reason one might worry about the dispersion relation is that IR divergences would cause a singularity at zero energy, which prevent us from pulling the contour around the origin to complete the argument.  Since the amplitude vanishes in perturbation theory, there is no perturbative evidence for such an IR singularity. Furthermore, in perturbative string theory, where such power law corrections to the $SU(8)$ violating amplitude certainly exist, and which has the same low energy spectrum and interactions, there is again no evidence than the amplitude is non-analytic at zero.  

\end{itemize}

To summarize: The $SU(8)$ violating five point amplitude, which follows from the existence of black holes of small charge above the production threshold for such black holes, has an analytic continuation which has a power law tale, and contradicts the existence of an $SU(8)$ invariant method for constructing a finite unitary S-matrix to all orders in a power series in energy divided by the Planck mass.  

There is a logical possibility that some of the methods in \cite{interest} might produce a finite result to all orders in perturbation theory, and provide evidence for a model of quantum gravity, which did not contain the BPS states, that are the focus of this article.  While I cannot provide a string free refutation of this claim, it is worth recalling how the ``UV" divergence of the pure SUGRA amplitudes would be interpreted in the string based computations of \cite{greenetal}. Their computations try to define SUGRA as a certain limit of Type II string amplitudes.   All such limits are equivalent
to weakly coupled string theory, or 11D SUGRA, on a large torus,
so in this method of calculation, the UV divergence must be interpreted
as an IR divergence coming from non-perturbative states whose mass goes to zero in this limit.

There is one other bit of information that emerged from thinking about the consequences of a finite $N=8$ SUGRA, which applies to any putatively finite but perturbatively non-renormalizable theory\cite{classicalization}.  In such a theory the expansion parameter is energy divided by some scale.  On the other hand, simple diagram counting shows that field theoretic perturbation series are divergent and lead to essential singularities in the coupling\footnote{Modern on shell methods for computing the S-matrix automatically combine together many diagrams and exhibit cancelations due to gauge invariance explicitly.  However, the $n\!$ ways of combining the on shell legs in an $n$-particle unitarity cut, mirrors the source of the factorial growth of diagrams. The path integral derivation of the degree of divergence of the perturbation series applies equally well to on shell amplitudes.} In such a theory, this would mean essential singularities at zero energy.  In the theories discussed in \cite{classicalization}, nothing prevents us from giving mass to all the fields in the Lagrangian.  Rigorous theorems of S-matrix theory then preclude the existence of essential singularities, either at the origin or at a finite mass.  The point is that in a theory with a mass gap, the number of states at finite energy is bounded and all singularities are multi-particle cuts, for a finite number of particles.  Thus, we can refute the claims that any such theory is finite, unless some principle prevents us from giving mass to the particles.  Of course, in $N=8$ SUGRA, there is such a principle, so this argument is not applicable.  That is why we went through the more elaborate discussion outlined above.

One might want to preserve the claim to a finite $N=8$ SUGRA for a bit longer by insisting that the finite theory does not have {\it any} charged states in it.   The classical Lagrangian has solutions corresponding to a pair of oppositely charged BPS black holes moving away from each other, with arbitrary initial velocity, from an arbitrary initial separation.  For some values of these parameters the pair will separate and reach infinity, while for others they will collapse and form a neutral black hole.  The usual rules of crossing symmetry suggest that charged particles pairs will be produced in the Hawking radiation of neutral black holes, so it is likely that one would have to
forbid the existence of ordinary Schwarzschild black holes as well.  Indeed,
we've already argued that the covariant entropy bound implies that processes involving the formation and decay of neutral black holes cannot preserve $SU(8)$ symmetry.  

 The other option is would be to declare that the theory contained no neutral black holes either, but then its claim to be a theory of gravitation would be empty.  The rest of this paper will be devoted to a more detailed discussion of the $SU(8)$ violating five point amplitude.

\section{Production of Soliton States}

The Lagrangian of $N=8$ SUGRA contains $28$ $U(1)$ gauge bosons. Among the solutions of the theory are BPS dyonic black holes with arbitrary continuous values of the electric and magnetic charges.  The theory has a classical $E(7,7)$ symmetry, which is partly realized by duality rotations in the $56$ dimensional charge space.  Dirac quantization restricts the space of charges to an integer lattice, which is invariant only under an integer subgroup $E(7,7;Z)$.  The $70$ scalar fields of the theory lie in the coset
$E(7,7)/SU(8)$, so only the $SU(8)$ symmetry acts linearly on scattering amplitudes for excitations around a particular point in moduli space. The coset transformations instead give rise to soft scalar low energy theorems for these amplitudes.  Recently, these symmetry constraints have been shown to guarantee the finiteness of all amplitudes that have so far been computed in this theory, but fail to rule out a divergence at $7$ loops.

The entropy of these BPS black holes is computed using the Bekenstein-Hawking law, which determines the large entropy asymptotics.   It is given by
$$S = \pi \sqrt{J_4},$$ where $J_4$ is an $E(7,7)$ invariant, quartic in the charges.  When the charges are quantized it is invariant only under the integer subgroup.  These are extremal black holes, so the charge also determines the mass, which is not invariant under $SU(8)$.  

The charges live on a Lorentzian lattice and the invariant $J_4$ can be positive, negative or zero.  The states with zero $J_4$ go to zero mass at certain boundaries of moduli space.  The corresponding classical solutions are singular, because they have horizons of zero area.  The authors of reference\cite{bfk} argued that these singular solutions might not be real quantum states of the theory in question.  While I do not agree with those arguments, the main discussion of the present paper will be restricted to states with non-zero $J_4$ .

The realization of $N=8$ SUGRA as the low energy effective field theory of Type II string theory compactified on a $6-$ torus of stringy size, sheds light on these states.   Some of the BPS sectors are realized in terms of Kaluza-Klein momenta, and wrapping numbers of elementary strings and D-branes on cycles of the torus\cite{balasubramanian}.  As shown by Witten\cite{witten}, all the boundaries of moduli space are dual (via $E(7,7;Z)$ rotations) to regimes in which weakly coupled Type II string theory, or 11 dimensional SUGRA, are compactified on a torus of growing size.  The $J_4 = 0$ states are precisely those which become massless particles at one of these boundaries.  This explains the vanishing of the area term in their entropy formula (the entropy is constant on moduli space).

When the string coupling is weak, and all geometrical moduli are of order string scale or larger, we can do reliable perturbative calculations of amplitudes involving virtual charged BPS states.  These lead to $SU(8)$ violation.  In string perturbation theory, only states with $J_4 = 0$ contribute
to the amplitudes.  We can also do reliable weak coupling calculations of 
amplitudes involving loops of wrapped D-branes with $J_4 \neq 0$.  These are T-dual to D-instanton calculations and are of order $e^{-\frac{c}{g_S}}$.  We will see below that this exponential suppression should be interpreted as 
an exponential of the linear size of the wrapped D-brane state, divided by its Compton wavelength.  Here we note only that this is a mathematical fact.  The size of a D-brane can be measured by scattering gravitons from it, and this leads to a size of order string scale\cite{Klebanov}.  The Compton wavelength is smaller than this by a factor $g_S$.\footnote{Note that this in no way contradicts the observation of \cite{DKPS} that D-branes in {\it e.g.} Type IIA, have a core of order the eleven dimensional Planck length, which is revealed in D-brane-D-brane scattering at low velocity.}   

Note further, that, if we imagine extrapolating to $g_s \sim 1$, there is every indication that the contribution of all of these BPS states to $SU(8)$ violating amplitudes will become of order one in eleven dimensional Planck units.
Elvang and Kiermaier\cite{helvkeir} have computed the lowest dimension $SU(8)$ violating operator in the effective field theory for the SUGRA multiplet, to lowest order in Type II string perturbation theory.   It has the form
$$  e^{- 6\phi} R^4 ,  $$  where $\phi$ is the dimensionless dilaton field. 
This term comes from explicit violation of $SU(8)$ in tree level string diagrams involving massive non-BPS string states, but it is of the same form as one would expect for the leading low energy contribution from loops of 
BPS states whose masses violate $SU(8)$.  In different regions of the moduli space $E(7,7)/SU(8)$, the field $\phi$ would be replaced by other members of the ${\bf 70}$ multiplet of Goldstone bosons.
Naively extrapolating to the regime $g_S \sim 1$ one finds that this term is scaled by the 11 D Planck mass, with a coefficient of order 1.  Of course, in this regime, the calculation of \cite{helvkeir} is not valid, but since there is no small dimensionless parameter,
it is hard to see how the term could be significantly smaller than the extrapolation of the perturbative calculation.   

Some of the advocates of a finite $N=8$ theory emphasize that their theory may have nothing to do with string theory and may be a new paradigm for the approach to quantum gravity as a quantum field theory. Therefore, I must emphasize that the above remarks about string theory are meant to be suggestive only.  Type II string theory with $g_S \sim 1$ and a torus of 11 dimensional Planck size is a finite theory with a single   parameter, with dimensions of mass.  Like the would be finite $N=8$ theory,
it has BPS states, whose mass formula violates the continuous symmetry of the leading term in the low energy effective Lagrangian.  Unlike that hypothetical theory, the violation shows up in the low energy power series expansion of the scattering amplitudes of the particles in the SUGRA multiplet.  What I will try to demonstrate below is that rather general principles imply that {\it any} sensible theory containing these BPS states will behave like string theory {\it does} and not like the hypothetical finite theory {\it is supposed to}.

\section{Brief outline of finite calculations}

This tiny section is {\it not} meant to be even a cursory introduction to the remarkable set of methods that have been developed to calculate the S-matrix in $N=8$ SUGRA and similar theories.   These include spinor helicity methods, tree level recursion relations based on analyticity in the complex momentum plane, and standard as well as novel unitarity cut techniques for calculating loop amplitudes.  Using dimensional regularization, some authors have even found a way to get around the traditional dispersion techniques for calculating the real parts of amplitudes.  The point is that in $4 -\epsilon$ dimensions, even ``polynomial" contributions to amplitudes have an imaginary part, as a consequence of the irrational scaling exponents.  Although the dimensionally continued amplitudes have no Hilbert space interpretation, the do satisfy the cutting rules, which are the perturbative expression of unitarity.

The important observation is that, as long as these methods give finite results, those results will be $SU(8)$ covariant\footnote{But see the remarks in Footnote 2.}.  The tree amplitudes are all $SU(8)$ covariant.   The loops are computed from the trees by expressing their discontinuities in terms of lower order amplitudes, and so $SU(8)$ covariance follows by induction, as long as the procedure gives finite and unambiguous answers at each step.  In particular, the five point amplitude that will be the focus of our attention, which transforms as a $70$, will be identically zero to all orders in the momentum expansion.  We will show that if we accept the existence of BPS states, with charges of order $1$, which are produced above threshold with amplitudes of order $1$, that the all orders vanishing of this amplitude contradicts some very general principles of S-matrix theory.

\section{Dispersion relations in models of quantum gravity}

The S-matrix elements for the scattering amplitude of two gravitons to two gravitons plus a scalar are conveniently classified according to the helicities of the incoming and outgoing gravitons.  Each helicity amplitude has a fixed dependence on components of the momenta, multiplied by unknown scalar functions. The scalar amplitudes in the $2\rightarrow 3$ S-matrix element depend on the square of the center of mass energy $s_{12} \equiv s$, and the other invariants $s_{ij} \equiv p_i \cdot p_j , $ (with $i \neq j$), subject to the constraints $\sum_j s_{ij} = 0$.  There are $5$ independent invariants.
We will write dispersion relations for these scalar functions in the variable $s$, with the other independent invariants held fixed.   

Feynman diagram calculations in low energy field theory, as well as perturbative string theory calculations indicate that the (somewhat vague) holomorphy properties of scattering amplitudes, which are postulated in analytic S-matrix theory, remain true in models of quantum gravity in asymptotically flat space.  Amplitudes are holomorphic functions of the invariants, with isolated singularities corresponding to the production of real on-shell physical particle states.  

Dispersion relations employ this assumption of holomorphy with an assumption that amplitudes are polynomially bounded in the entire complex $s$ plane.  This assumption is valid in low energy effective field theory, but when the perturbation parameter has dimensions of length to a positive power, the order of the polynomial increases with the order of perturbation theory.  

In fact, at fixed {\it impact parameter}, simple considerations of black hole physics suggest that the amplitudes fall off exponentially at large $s$.  The argument is simple.   As $s$ increases, the colliding particles in the center of mass frame will have a gravitational field, as seen by a distant observer
$$ds^2 = - dt^2 (1 - \frac{2\sqrt{s} G_N}{r}) + \frac{dr^2}{(1 - \frac{2\sqrt{s} G_N}{r})} + r^2 d\Omega^2 .$$  When the impact parameter is less than $2 \sqrt{s} G_N$, the colliding particles are inside the horizon of the black hole.  It is extremely plausible (in general relativity, this is called The Hoop Conjecture) that in such a kinematic regime, particle collisions lead to black hole formation, with probability that rapidly approaches $1$ as $s$ goes to infinity with fixed impact parameter.

When $s$ is far above the threshold for black hole production at some fixed impact parameter, we can treat the black hole by semi-classical techniques.  In particular, Hawking's calculation of black hole decay tells us that the probability of producing a particular final state, with a small number of particles carrying all of the outgoing energy, is {\it exponentially} suppressed.
The argument is simple counting.  The black hole decay is thermal, and the black hole entropy increases linearly with $s$.  

Giddings and Porto\cite{giddport} have argued that for fixed $s_{ij}$ this behavior implies that the standard assumption of polynomial boundedness in the complex $s$ plane is still valid.  The amplitudes at fixed kinematic invariants are integrals over {\it all} impact parameters.  Since we have argued that for small enough impact parameter, in a range growing like $\sqrt{s}$ , the amplitudes fall off exponentially, the high energy behavior at fixed $s_{ij}$ is dominated by large impact parameters, where it may be estimated by the eikonal approximation.  The eikonal behavior obeys the Cerulus-Martin lower bound on the real axis, and there is every indication that the amplitude is power law bounded in the entire complex $s$ plane, for fixed values of the other $s_{ij}$, and should satisfy a dispersion relation with a finite number of subtractions.  

For the $2 \rightarrow 2$ amplitude in ${\cal N} = 8$ SUGRA, we expect the right and left hand cuts, which contribute to the dispersion integrand, to meet at the origin.  However, for the $SU(8)$ violating 5-point amplitude at hand, one would imagine that the lowest contributing threshold would correspond to 
the production of pairs of the lightest charged BPS states, whose masses are of order the Planck scale\footnote{Here I'm assuming an hypothetical theory in which the leading source of $SU(8)$ violation comes from BPS states with $J_4 \neq 0$.  In string perturbation theory, breaking to a continuous subgroup occurs already with exchange of massive string modes at tree level.  Loops of $J_4 = 0$ states break $SU(8)$ to a discrete subgroup, while the states with $J_4 \neq 0$ are non-perturbative in $g_S$ because their size is much larger than their Compton wavelength. For string theory with $g_S$ and radii in string units all of order one, or any model whose only parameter is the 4D Planck mass, the threshold for $SU(8)$ violating intermediate states will be given by this scale.}.  The amplitudes will have IR divergences, corresponding to emission of soft gravitons from incoming and outgoing particles, as they accelerate, but these will not lead to a lowering of the threshold for real production of particles whose mass violates $SU(8)$.  Both the spectrum and perturbative interactions of the SUGRA multiplet preserve $SU(8)$.

We have thus argued that $SU(8)$ violating helicity amplitudes may be written as

$$ A^h (s) = K^h (s, s_{ij}) [P(s) + \int_{s_0}^{\infty}\ \frac{dx}{(x - s)^N} \rho (x) ],$$ where $N$ is an integer we have not determined, and $K^h$ is a known kinematic helicity factor.  $P(s)$ is a polynomial, and $\rho$ is the sum of the discontinuities across the right and left hand cuts.  $s_0$ is the squared mass of the lightest BPS state contributing to the discontinuity.  It is of order $m_P^2$.  $P$ and $\rho$ depend implicitly on the other $s_{ij}$, which we take to be fixed at non-zero values $\ll m_P^2$.  

The low energy expansion of this amplitude involves the moments
$$M_k = \int_{1}^{\infty} \rho (y s_0 ) y^{- (N + k)} = \int_{0}^{1}\ dz \sigma (z) z^{N+k - 2},$$ where $\sigma (z) \equiv \rho(\frac{s_0}{z} ), $ and the integrals are all power law convergent at $z =0 $.  

The assumption of a convergent perturbation expansion, constructed in terms of recursion relations and perturbative unitarity, implies that $A^h (s)$ vanishes to all orders in $\frac{s}{m_P^2}$.  This in turn implies that all moments above some $k_{max}$ vanish.  The function $$f(y) = \rho (y s_0 ) y^{-N} ,$$ is square integrable with respect to the measure $d\mu = \frac{dy}{y^{k_{max}}}$.  Indeed, it is finite at every $y$, because it is a sum of discontinuities across multiparticle cuts, and the integral $$\int_1^{\infty}\ \frac{dy}{y^{2N + k_{max}}}\ |\rho(y s_0) |^2 , $$ is convergent at infinity because of the convergence of the original dispersion relation.  

The vanishing of the higher moments is equivalent to the statement that $f(y)$ is orthogonal to all of the functions $y^{-k}$ for non-negative $k$.  These functions are square integrable and linearly independent.  By Gram-Schmidt orthogonalization, we can use them to construct an orthonormal basis.  Consequently $f(y) = 0$.  This contradicts the explicit finite contribution to $f(y)$ slightly above the production threshold for the lightest BPS dyon-anti-dyon pair.

\subsection{Production cross sections for BPS monopoles}

It is sometimes argued that the BPS states are ``non-perturbative", and should not show up in the perturbation expansion.  This response is too superficial, and a proper understanding of why soliton production is suppressed in conventional quantum field theory shows that there is no similar suppression of production of BPS states of small charge.

Indeed, in conventional field theory, solitons have masses of order $\frac{m}{g^2}$, where $m$ is the mass of perturbative excitations, and $g$ the dimensionless coupling.  Energy denominator suppression would indicate that they contribute to scattering of perturbative excitations at finite orders in perturbation theory, but this is not true.  The example of magnetic monopoles in the $SO(3)$ Georgi-Glashow model, is typical.  Consider the contribution of monopole production to the total cross section for annihilation of two perturbative charged excitations, at lowest order in the electro-magnetic coupling.  It is determined by the vacuum to monopole-anti-monopole matrix element of the electromagnetic current:
$$\langle 0 | J^{\mu} (0) | M \bar{M} \rangle .$$  By crossing symmetry, this is the analytic continuation of the monopole form factor
$$ \langle M (0) | J^{0} | M (q) \rangle ,$$ to time-like $q$, $\geq$ twice the monopole mass.   The usual rules of semi-classical physics, tell us that the monopole form factor in the monopole rest frame is simply the classical electromagnetic current of the monopole field configuration.  The current is a smooth function  $j ( x m )$, where $m$ is the mass of the charged vector boson.  The Riemann-Lebesgue lemma, tells us that at large $q$ the function falls off faster than any power.  It is typically of order
$$e^{- a\frac{| q l}{m}}, $$ where the constant $a$ is of order one and represents the closest singularity of $j (xm ) $ to the real axis in the complex $xm$ plane.  This leads to the estimate 
$$e^{ - \frac{2a}{g^2}},$$ for the monopole production amplitude.  

It is clear from this calculation that the relevant parameter is the ratio between the physical size of the soliton, and its Compton wave-length.  Non-perturbative suppression of soliton production means that the amplitudes are exponentially small in this ratio. Although I've given this argument in the context of the Georgi-Glashow model, this explanation of the non-perturbative suppression of soliton production amplitudes above threshold is universal in all quantum field theories.

The lesson for production of charged BPS states in ${\cal N} = 8$ SUGRA is obvious.  The scale of variation of the fields is the Schwarzschild radius, which goes like the square root of the $E_{7,7}$ invariant, $J_4$, that determines the entropy.  Indeed, this is the only dimensionless parameter, which characterizes these solutions.  We conclude that when $J_4$ is of order one, there is no suppression of BPS production amplitudes.

One might attempt to argue that we only have reliable evidence for the existence of these BPS states when $J_4$ is large and the semi-classical description of the states is valid.  However, this is enough to prove the existence of states of large charge of opposite sign and almost equal magnitude.  Cluster decomposition and charge conservation assure us that these can annihilate to states of small charge, which should then Hawking radiate down to the BPS state of small charge.  Previously this argument was made only for states of $J_4 = 0$, and\footnote{In my opinion, specious,} arguments were given that those states might not exist.  We've now seen that the argument holds just as well for states with small non-zero $J_4$, and that the existence of such states follows from the existence of
non-singular BPS black holes.

We thus have arguments based on very general principles, that $SU(8)$ violation should occur at finite orders in the expansion in energy over the Planck scale.  This manifestly contradicts the claim that the techniques for constructing finite amplitudes will construct the correct amplitudes to all orders, since those techniques preserve $SU(8)$.  With more work, one could perhaps give an estimate of the order in the loop expansion at which divergences occur.
Elvang and Kiermaier\cite{helvkeir}, have found the leading low energy contribution to $SU(8)$ violating amplitudes in perturbative string theory.  These come from non-BPS massive string modes, at tree level in the $g_S$ expansion\footnote{Which is not the same as the ${\cal N} = 8$ loop expansion.}.  By unitarity, these give higher order contributions to the $2 \rightarrow 2$ super-graviton amplitude, which cannot be captured by the methods that give finite perturbative amplitudes.  Perturbative finiteness should fail before these contributions become manifest.  Henriette Elvang (private communication) has pointed out that if we assume the $SU(8)$ violating $5$ point function appears at the order in the energy expansion indicated by the lowest dimension operator that appears in string theory, then the induced unitary correction to $2 \rightarrow 2$ graviton scattering first shows up at $8$ loops in the SUGRA loop expansion.  This is consistent with the conjecture that finiteness will fail at $7$ loops. Unfortunately, our general argument does not rule out the possibility that some finite order of the perturbation expansion of $SU(8)$ violating amplitudes vanishes.  String theory predicts definite orders, but if one takes the point of view that the hypothetical finite perturbation series defines a separate theory of quantum gravity, disjoint from string theory, one cannot counter the argument by citing string theory results.

Finally I'd like to note that, although I've concentrated on BPS states, where $SU(8)$ violation is manifest in the mass spectrum, ordinary Schwarzschild black holes are expected to lead to $SU(8)$ violation as well.  A continuous global symmetry like $SU(8)$ violates the Covariant Entropy Bound for black holes.  By dropping neutral particles with $SU(8)$ quantum numbers into a black hole, we can create an infinite number of distinct black holes with the same Schwarzschild field, but transforming in different $SU(8)$ representations.  Graviton-graviton scattering, at fixed impact parameter less than the Schwarzschild radius corresponding to the center of mass energy, is expected to be dominated by black hole production.  Once $\sqrt{s} > M_P$, this occurs at impact parameters larger than the Planck scale.  Black hole production and decay {\it must} violate $SU(8)$ to avoid the violation of the covariant entropy bound. 

As a consequence, it appears that there are only three logical responses to the arguments presented in this paper.

\begin{itemize}

\item The $SU(8)$ invariant perturbation series fails to be finite at some order.  This would indicate that ${\cal N} = 8$ SUGRA is just an effective field theory, which requires UV completion.  As occurs in string theory, the UV completion could violate $SU(8)$ and be consistent with the existence of charged BPS states.

\item The $SU(8)$ invariant perturbation expansion is finite to all orders.  If the considerations of this paper are correct, it cannot be the perturbation series of  a theory which contains charged BPS states, and probably not of one which contains any kind of black hole.  Despite the fact that it contains a graviton, one would feel uncomfortable calling such a model a quantum theory of gravity.

\item One of the assumptions of this paper is incorrect, most likely the assumption that amplitudes are polynomially bounded in the cut complex energy plane.  The results of Giddings and Porto make it plausible that gravitational amplitudes obey such a bound, but by their nature they cannot constitute a proof of that fact.

\end{itemize}

\vfill\eject

\begin{center}
{\bf Acknowledgements}
\end{center}

I would like to thank Dan Freedman, Andy Strominger, Greg Moore, Renata Kallosh, Lance Dixon, Stanley Deser, Kelly Stelle, Michael Green, Michael Kiermaier, Dan Harlow, Rafael Porto, Zvi Bern, and particularly Henriette Elvang, for numerous conversations which led me to an understanding of the issues discussed in this paper.  This work was supported in part by the Department of Energy.




\end{document}